\documentclass[pre,tighten]{revtex4}

\usepackage{amssymb,amsmath}

\usepackage{graphicx}

\begin{document}
\title{A note on the violation of the Einstein relation in a driven moderately dense granular gas}
\author{Vicente Garz\'o}
\email{vicenteg@unex.es} \homepage{http://www.unex.es/eweb/fisteor/vicente/} \affiliation{Departamento de
F\'{\i}sica, Universidad de Extremadura, E-06071 Badajoz, Spain}
\begin{abstract}
The Einstein relation for a driven moderately dense granular gas in $d$-dimensions is
analyzed in the context of the Enskog kinetic equation. The Enskog equation neglects
velocity correlations but retains spatial correlations arising from volume exclusion
effects. As expected, there is a breakdown of the Einstein relation
$\epsilon=D/(T_0\mu)\neq 1$ relating diffusion $D$ and mobility $\mu$, $T_0$ being the
temperature of the impurity. The kinetic theory results also show that the violation of
the Einstein relation is only due to the strong non-Maxwellian behavior of the
reference state of the impurity particles. The deviation of $\epsilon$ from unity
becomes more significant as the solid volume fraction and the inelasticity increase,
especially when the system is driven by the action of a Gaussian thermostat. This
conclusion qualitatively agrees with some recent simulations of dense gases [Puglisi
{\em et al.}, 2007 {\em J. Stat. Mech.} P08016], although the deviations observed in
computer simulations are more important than those obtained here from the Enskog
kinetic theory. Possible reasons for the quantitative discrepancies between theory and
simulations are discussed.
\end{abstract}
\draft
\date{\today}
\maketitle

\section{Introduction}
\label{sec1} The generalization of the fluctuation-response relation to non-equilibrium systems has received a
considerable attention in the past few years. In this context, granular matter can be considered as a good
example of a system that inherently is in a non-equilibrium state. Granular systems are constituted by
macroscopic grains that collide inelastically so that the total energy decreases with time. On the other hand, a
non-equilibrium steady state (NESS) is reached when the system is {\em heated} by the action of an external
driving force (thermostat) that does work to compensate for the collisional loss of energy. In these conditions,
some attempts to formulate a fluctuation-response theorem based on the introduction of an effective temperature
have been carried out \cite{PBL02,SD04,SBL06,BSL08}. However, a complete analysis of the validity of the theorem
requires the knowledge of the full dependence of the response and correlation functions on frequency $\omega$
\cite{L89}. Given that this dependence is quite difficult to evaluate in general, the corresponding limit
$\omega \to 0$ is usually considered. In this limit, the classical relation between the coefficients of
diffusion $D$ (autocorrelation function) and mobility $\mu$ (linear response) is known as the Einstein relation.

The Einstein relation for heated granular fluids has been widely analyzed recently.
First, some computer simulation results for dilute systems \cite{BLP04} have shown the
validity of the Einstein relation ($\epsilon=D/T_0 \mu=1$) in NESS when the temperature
of the bath $T$ is replaced by the temperature of the impurity $T_0$. This has an
interesting consequence in the case of mixtures (where the different species have
different temperatures \cite{GD99,nonequip,exp}) since a linear response experiment on
a {\em massive} intruder or tracer particle to obtain a temperature measurement yields
the temperature of the intruder and not the temperature of the surrounding gas. On the
other hand, from an analytical point of view, kinetic theory calculations  based on the
Boltzmann equation have shown the violation of the Einstein relation ($\epsilon \neq
1$) in the free cooling case \cite{DG01} as well as for driven granular gases
\cite{G04}. These deviations are in general very small in the driven case (less than
$1\%$ when the system is driven by a stochastic thermostat) and are related to
non-Gaussian properties of the distribution function of the impurities. This is the
reason why such deviations cannot be detected in computer simulations of very dilute
gases.

However, a recent computer simulation study of Puglisi {\em et al.} \cite{PBV07} at high densities has provided
evidence that the origin of the violation of the Einstein formula is mainly due to spatial and velocity
correlations between the particles that are about to collide rather than the deviation from the
Maxwell-Boltzmann statistics. These correlations increase as excluded volume and energy dissipation occurring in
collisions are increased. The simulation results obtained by Puglisi {\em et al.} \cite{PBV07} motivate the
present paper and, as in the case of a dilute gas \cite{G04}, kinetic theory tools will be used to analyze the
effect of density on the possible violation of the Einstein relation. For a moderately dense gas, the Enskog
kinetic equation for inelastic hard spheres \cite{BDS97} can be considered as an accurate and practical
generalization of the Boltzmann equation. As in the case of elastic collisions, the Enskog equation takes into
account spatial correlations through the pair correlation function but neglects velocity correlations (molecular
chaos assumption) \cite{F72}. Although the latter assumption has been shown to fail for inelastic collisions as
the density increases \cite{ML98,SM01,PTNE02}, there is substantial evidence in the literature for the validity
of the Enskog theory for densities outside the Boltzmann limit (moderate densities) and values of dissipation
beyond the quasielastic limit. This evidence is supported by the good agreement found at the level of
macroscopic properties (such as transport coefficients) between the Enskog theory \cite{GD99b,L05,GDH07} and
simulation \cite{L01,LBD02,DHGD02,MGAL06,LLC07} and experimental \cite{YHCMW02,HYCMW04} results. In this
context, one can conclude that the Enskog equation provides a unique basis for the description of dynamics
across a wide range of densities, length scales, and degrees of dissipation. No other theory with such
generality exists.

\section{Description of the problem}
\label{sec2}

Let us consider a granular gas composed by smooth inelastic disks ($d=2$) or spheres
($d=3$) of mass $m$, diameter $\sigma$, and interparticle coefficient of restitution
$\alpha$  in a {\em homogeneous} state. At moderate densities, we assume that the
velocity distribution function $f({\bf v})$ obeys the Enskog kinetic equation
\cite{BDS97}. Due to dissipation in collisions, the gas cools down unless a mechanism
of energy input is externally introduced to compensate for collisional cooling. In
experiments the energy is typically injected through the boundaries yielding an
inhomogeneous steady state. To avoid the complication of strong temperature
heterogeneities, it is usual to consider the action of  homogeneous external (driving)
forces acting locally on each particle. These forces are called {\em thermostats} and
depend on the state of the system. In this situation, the steady-state Enskog equation
reads
\begin{equation}
\label{1} {\cal F}f({\bf v})=\chi J[{\bf v}|f,f],
\end{equation}
where $J[{\bf v}|f,f]$ is the inelastic Boltzmann collision operator, $\chi$ denotes the equilibrium
configurational pair correlation function evaluated at contact, and ${\cal F}$ is an operator representing the
effect of the external force. Two types of external forces (thermostats) are usually considered: (a) a
deterministic force proportional to the particle velocity (Gaussian thermostat), and (b) a white noise external
force (stochastic thermostat). The use of these kinds of thermostats has attracted the attention of granular
community in the past years to study different problems. In the case of the Gaussian thermostat, ${\cal F}$ has
the form \cite{EM90,H91,MS00}
\begin{equation}
\label{4} {\cal F}f({\bf v})=\frac{1}{2}\zeta \frac{\partial}{\partial {\bf v}}\cdot \left[{\bf v}f({\bf
v})\right],
\end{equation}
where $\zeta$ is the cooling rate due to collisions. In the case of the stochastic thermostat, the operator
${\cal F}$ has the Fokker-Planck form \cite{NE98}
\begin{equation}
\label{7} {\cal F}f({\bf v})=-\frac{1}{2}\frac{T}{m}\zeta \left(\frac{\partial}{\partial {\bf v}}\right)^2f({\bf
v}).
\end{equation}
The exact solution to the Enskog equation (\ref{1}) is not known, although a good
approximation for $f$ in the region of thermal velocities can be obtained from an
expansion in Sonine polynomials. For practical purposes, one selects a finite number of
terms in the expansion. In the leading order $f({\bf v})$ is given by
\begin{equation}
\label{7.1} f({\bf v})\to n\pi^{-d/2}v_{\text{th}}^{-d}e^{-
v^{*2}}\left[1+\frac{c}{4}\left(v^{*4}-(d+2) v^{*2}+\frac{d(d+2)}{4}\right)\right],
\end{equation}
where ${\bf v}^*={\bf v}/v_{\text{th}}$, $v_{\text{th}}=\sqrt{2T/m}$ being the thermal
velocity. Moreover, $c$ is the fourth cumulant of the velocity distribution function
$f$ defined as
\begin{equation}
\label{8.1} c=\frac{8}{d(d+2)}\left(\frac{m^2}{4nT^2}\int\;d{\bf v}\; v^4
f-\frac{d(d+2)}{4}\right).
\end{equation}
In the approximation (\ref{7.1}), cumulants of higher order have been neglected.
Inserting Eq.\ (\ref{7.1}) into the Enskog equation and neglecting nonlinear terms in
$c$, one gets the following expression for the cooling rate $\zeta$ \cite{NE98}
\begin{equation}
\label{8} \zeta=\frac{\sqrt{2}\pi^{(d-1)/2}}{d\Gamma\left(\frac{d}{2}\right)}\chi n
\sigma^{d-1} v_{\text{th}} (1-\alpha^2)\left(1+\frac{3}{32}c\right).
\end{equation}
The value of $c$ depends on the thermostat used. In the case of the Gaussian
thermostat, $c$ is approximately given by \cite{NE98}
\begin{equation}
\label{9} c(\alpha)=\frac{32(1-\alpha)(1-2\alpha^2)}{9+24d-(41-d)\alpha+30\alpha^2(1-\alpha)},
\end{equation}
while
\begin{equation}
\label{10} c(\alpha)=\frac{32(1-\alpha)(1-2\alpha^2)}{73+56d-3(35+8d)\alpha+30\alpha^2(1-\alpha)}
\end{equation}
for the stochastic thermostat. It is interesting to remark that in the homogeneous
problem the results obtained with the Gaussian thermostat are completely equivalent to
those derived in the free cooling case when one scales the particle velocity with
respect to the thermal velocity $v_{\text{th}}$ \cite{MS00}. In addition, although the
expressions (\ref{8})--(\ref{10}) have been derived by neglecting nonlinear terms in
the coefficient $c$, the estimates (\ref{9}) and (\ref{10}) present quite a good
agreement with Monte Carlo simulations of the Boltzmann equation \cite{MS00,BMC96} for
moderate values of dissipation (say for instance, $\alpha\gtrsim 0.5$). However, more
recent results \cite{BP06,NBSG07} for the homogeneous (undriven) cooling state have
shown that for very large inelasticity ($\alpha\lesssim 0.5$), the higher-order
cumulants may not be neglected since they can be of the same order of magnitude as $c$.
The breakdown of the Sonine polynomial expansion is caused by the increasing impact of
the overpopulated high-energy tail of the velocity distribution $f$ \cite{PBF06}.

We assume now that a few impurities or tracer particles of mass $m_0$ and diameter
$\sigma_0$ are added to the system. Given that their concentration is very small, the
state of the gas is not affected by the presence of impurities. As a consequence, the
velocity distribution function $f$ of the gas still verifies the (homogeneous) Enskog
equation (\ref{1}). Moreover, one can also neglect collisions among impurities
themselves versus the impurity-gas collisions, which are characterized by the
coefficient of restitution $\alpha_0$. Diffusion of impurities is generated by a weak
concentration gradient $\nabla n_0$ and/or a weak external field ${\bf E}$ (e.g.
gravity or an electric field) acting only on the impurity particles.  Under these
conditions, the velocity distribution function $f_0({\bf r}, {\bf v},t)$ of impurities
verifies the Enskog-Lorentz equation
\begin{equation}
\partial_t f_0+{\bf v}\cdot \nabla f_0+\frac{{\bf E}}{m_0}\cdot \frac{\partial}{\partial {\bf v}}f_0+{\cal F}f_0
=\chi_0 J[{\bf v}|f_0,f], \label{11}
\end{equation}
where $J\left[f_0,f\right]$ is the (inelastic) Boltzmann-Lorentz collision operator and $\chi_0$ represents the
equilibrium pair correlation function for impurity-fluid pairs at contact. Given that the gas is in a
homogeneous state, it follows that $\chi_0$ is uniform. At a kinetic level, an interesting quantity is the
partial temperature of impurities $T_0$. It is defined as
\begin{equation}
\label{14} \frac{d}{2}n_0T_0=\int\, d{\bf v}\, \frac{m_0}{2}v^2 f_0({\bf v}),
\end{equation}
where $n_0$ is the number density of impurities. The corresponding cooling rate $\zeta_0$ associated with the
partial temperature $T_0$ of impurities is defined as
\begin{equation}
\label{15} \zeta_0=-\frac{\chi_0}{dn_0T_0}\int d{\bf v} m_0v^2 J[{\bf v}|f_0,f].
\end{equation}
\begin{figure}
\includegraphics[width=0.4 \columnwidth,angle=0]{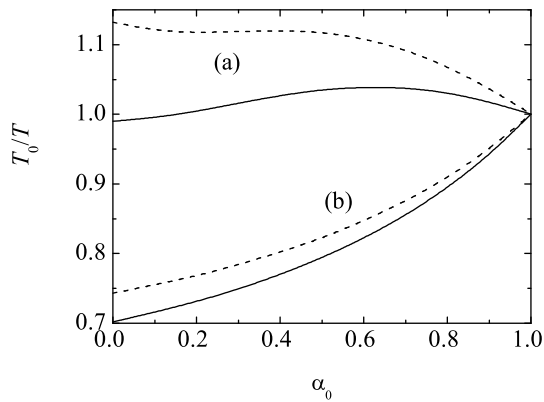}
\caption{Plot of the temperature ratio  $T_0/T$ versus the coefficient of restitution $\alpha=\alpha_0$ for
$d=3$ in the case  $m_0/m=\sigma_0/\sigma=0.5$ for the Gaussian thermostat (a) and the stochastic thermostat
(b). The solid lines correspond to a dilute gas ($\phi=0$) while the dashed lines refer to a moderately dense
gas ($\phi=0.2$).\label{fig1}}
\end{figure}
\begin{figure}
\includegraphics[width=0.4 \columnwidth,angle=0]{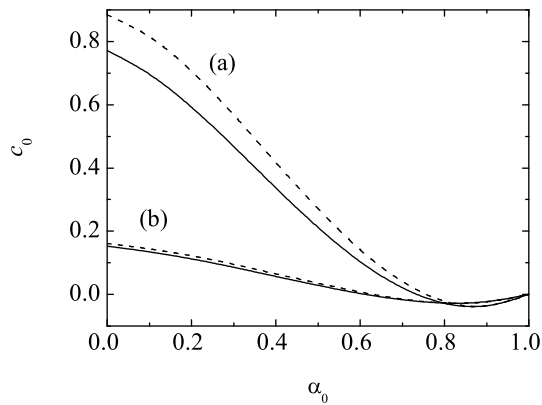}
\caption{Plot of the coefficient $c_0$ versus the coefficient of restitution $\alpha=\alpha_0$ for $d=3$ in the
case $m_0/m=\sigma_0/\sigma=0.5$ for the Gaussian thermostat (a) and the stochastic thermostat (b). The solid
lines correspond to a dilute gas ($\phi=0$) while the dashed lines refer to a moderately dense gas ($\phi=0.2$).
\label{fig2}}
\end{figure}

In the absence of diffusion (homogeneous steady state), Eq.\ (\ref{11}) becomes
\begin{equation}
\label{16} {\cal F}f_0=\chi_0J[{\bf v}|f_0,f].
\end{equation}
This equation has been widely analyzed by using both types of thermostats \cite{DHGD02,G04} for hard spheres
($d=3$). The results show that the temperatures of the gas ($T$) and impurities ($T_0$) are clearly different
and so the energy equipartition is broken down. In general, the temperature ratio $\gamma\equiv T_0/T$ presents
a complex dependence on the parameters of the problem. The condition for determining the temperature ratio
$\gamma$ is different for each type of thermostat. In the case of the Gaussian thermostat, the temperature ratio
is obtained by equating the cooling rates \cite{GM04,GD99}
\begin{equation}
\label{17} \zeta=\zeta_0,
\end{equation}
while for the stochastic thermostat $\gamma$ is obtained from the condition
\cite{DHGD02}
\begin{equation}
\label{18} \frac{\zeta T}{m}=\frac{\zeta_0 T_0}{m_0}.
\end{equation}
Requirements (\ref{17}) and (\ref{18}) lead to a different dependence of the temperature ratio $T_0/T$ on the
control parameters, namely, the mass ratio $m_0/m$, the size ratio $\sigma_0/\sigma$, the coefficients of
restitution $\alpha$ and $\alpha_0$, and the solid volume fraction
\begin{equation}
\phi\equiv \frac{\pi^{d/2}}{2^{d-1}d\Gamma\left(\frac{d}{2}\right)} n \sigma^d. \label{18.1}
\end{equation}
Apart from the temperature ratio, an interesting quantity is the fourth cumulant $c_0$.
It is defined as
\begin{equation}
\label{18.2} c_0=\frac{8}{d(d+2)}\left(\frac{m_0^2}{4n_0T_0^2}\int\;d{\bf v}\; v^4 f_0-\frac{d(d+2)}{4}\right).
\end{equation}
As in the case of the coefficient $c$, the cumulant $c_0$ measures the deviation of
$f_0$ from its Maxwellian form
\begin{equation}
\label{25.1} f_{0,M}({\bf v})=n_0\left(\frac{m_0}{2T_0}\right)^{d/2}\exp\left(-
\frac{m_0v^2}{2T_0}\right).
\end{equation}

In order to determine the coefficients $\zeta_0$ and $c_0$ one has to expand the
velocity distribution function $f_0$ in terms of the orthogonal Sonine polynomials.  As
in the case of the granular gas distribution $f$, a good estimate of $\zeta_0$ and
$c_0$ can be obtained from the leading Sonine approach to $f_0$:
\begin{equation}
\label{18.3} f_0({\bf v})\to n_0\pi^{-d/2}v_{\text{th}}^{-d}\theta^{d/2}e^{-\theta
v^{*2}}\left[1+\frac{c_0}{4}\left(\theta^2v^{*4}-(d+2)\theta
v^{*2}+\frac{d(d+2)}{4}\right)\right],
\end{equation}
where $\theta=m_0T/mT_0$ is the mean square velocity of the gas particles relative to
that of impurities. Expressions for $\zeta_0$ and $c_0$ have been derived in Appendix
\ref{appA} for an arbitrary number of dimensions $d$ by considering only linear terms
in $c$ and $c_0$. These expressions extend previous results derived in Ref.\ \cite{G04}
for hard spheres. Once $\zeta_0$ and $c_0$ are known, the temperature ratio can be
obtained from the constraints (\ref{17}) and (\ref{18}) for the Gaussian and stochastic
thermostats, respectively. To get this explicit dependence, the form of the pair
correlation functions $\chi$ and $\chi_0$ in terms of the size ratio $\sigma_0/\sigma$
and the solid volume fraction $\phi$ must be given. For a three-dimensional gas
($d=3$), a good approximation for these functions is \cite{GH72}
\begin{equation}
\label{19} \chi=\frac{1-\frac{1}{2}\phi}{(1-\phi)^3},
\end{equation}
\begin{equation}
\label{20}
\chi_0=\frac{1}{1-\phi}+\frac{3}{2}\frac{\sigma_0}{\overline{\sigma}}\frac{\phi}{(1-\phi)^2}+\frac{1}{2}
\left(\frac{\sigma_0}{\overline{\sigma}}\right)^2\frac{\phi^2}{(1-\phi)^3},
\end{equation}
where $\overline{\sigma}=(\sigma+\sigma_0)/2$. For a two-dimensional gas ($d=2$), $\chi$ and $\chi_0$ are
approximately given by \cite{JM87}
\begin{equation}
\label{21} \chi=\frac{1-\frac{7}{16}\phi}{(1-\phi)^2},
\end{equation}
\begin{equation}
\label{22} \chi_0=\frac{1}{1-\phi}+\frac{9}{16}\frac{\sigma_0}{\overline{\sigma}}\frac{\phi}{(1-\phi)^2}.
\end{equation}

Obviously, $\chi=\chi_0$ if $\sigma=\sigma_0$. Thus the temperature ratio and the
kurtosis $c_0$ become independent of density for equal--size particles. The dependence
of $T_0/T$ on the (common) coefficient of restitution $\alpha=\alpha_0$ is illustrated
in Fig.\ \ref{fig1} for $d=3$ in the case $m_0/m=\sigma_0/\sigma=0.5$ and for two
values of the solid volume fraction $\phi$. We consider the two types of thermostats
discussed before. There is an evident breakdown of the energy equipartition in both
thermostats, especially in the case of the stochastic driving force (\ref{7}). However,
the influence of density is more significant for the Gaussian thermostat than for the
stochastic one. The dependence of $c_0$ on $\alpha_0$ is plotted in Fig.\ \ref{fig2}
for the same cases as considered in Fig.\ \ref{fig1}. It is apparent that for both
thermostats the value of $c_0$ is quite small for not too inelastic systems. This means
that in this range of values of $\alpha_0$ the distribution $f_0$ of the homogeneous
state is quite close to a Maxwellian at the temperature of the impurity particle $T_0$.
However, the magnitude of $c_0$ increases significantly as the dissipation increases,
especially in the case of the Gaussian thermostat. This is a signal of the strong
non-Maxwellian behavior of the reference homogeneous state of impurities for quite
extreme values of inelasticity. As a consequence, cumulants of higher order than the
fourth cumulant $c_0$ should be considered to assess the deviation of $f_0$ from its
Maxwellian form $f_{0,M}$ for very small values of $\alpha$ and $\alpha_0$. Finally,
with respect to the influence of density, we observe that it is more relevant for the
Gaussian thermostat than for the stochastic thermostat, being practically negligible in
the latter case.
\begin{figure}
\includegraphics[width=0.4 \columnwidth,angle=0]{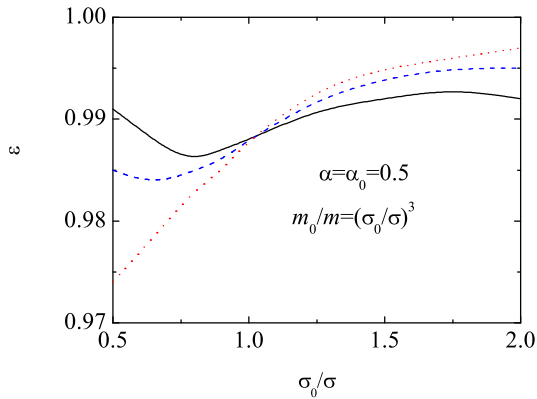}
\caption{(Color online) Plot of the Einstein ratio $\epsilon=D/T_0\mu$ versus the size
ratio $\sigma_0/\sigma$ in the case of the Gaussian thermostat for $d=3$ and
$\alpha=\alpha_0=0.5$ when the impurities have the same mass density as the gas
particles. Three different values of the solid volume fraction are considered: $\phi=0$
(solid line), $\phi=0.2$ (dashed line), and $\phi=0.4$ (dotted line). \label{fig3}}
\end{figure}

\section{The Einstein relation}
\label{sec3}

The Einstein ratio $\epsilon$ is defined as
\begin{equation}
\label{23.0} \epsilon=\frac{D}{T_0\mu},
\end{equation}
where $D$ and $\mu$ are the diffusion and mobility coefficients, respectively. If the Einstein relation would
hold, one would have $\epsilon=1$. I want here to analyze the influence of density on $\epsilon$. The transport
coefficients $D$ and $\mu$ can be determined by solving the (inelastic) Enskog-Lorentz equation (\ref{11}) by
means of the Chapman-Enskog method \cite{CC70}. In the first order of the expansion, the current of impurities
${\bf j}_0$ has the form \cite{G04}
\begin{equation}
\label{23} {\bf j}_0=-D \nabla \ln n_0+\mu {\bf E}.
\end{equation}
Given that $\chi$ and $\chi_0$ are uniform in this problem, it is evident that, when
properly scaled, the previous solution obtained in Ref.\ \cite{G04} for a dilute gas
can be directly translated to the Enskog equation by making the changes $J[f,f]\to \chi
J[f,f]$ and $J[f_0,f]\to \chi_0 J[f_0,f]$. Technical details on the calculation of $D$
and $\mu$ by means of the Chapman-Enskog expansion \cite{CC70} up to the second Sonine
approximation can be found in Ref.\ \cite{G04} for inelastic hard spheres ($d=3$). The
extension to an arbitrary number of dimensions is straighforward. Taking into account
these results, the dependence of the Einstein ratio $\epsilon$ on the parameter space
of the problem can be obtained. In the case of the Gaussian thermostat, the result is
\begin{equation}
\label{24} \epsilon=1-\frac{c_0}{2}\frac{\nu_2}{\nu_4-\frac{3}{2}\zeta},
\end{equation}
where the collision frequencies $\nu_2$ and $\nu_4$ are explicitly given in Appendix \ref{appA} for an arbitrary
number of dimensions $d$. The result for the case of the stochastic thermostat is
\begin{equation}
\label{25} \epsilon=1-\frac{c_0}{2}\frac{\nu_2}{\nu_4}.
\end{equation}
\begin{figure}
\includegraphics[width=0.4 \columnwidth,angle=0]{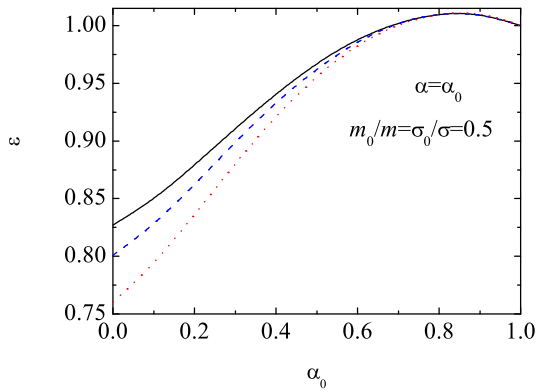}
\caption{(Color online) Plot of the Einstein ratio $\epsilon=D/T_0\mu$ versus the
coefficient of restitution $\alpha=\alpha_0$ in the case of the Gaussian thermostat for
$d=2$, $m_0/m=\sigma_0/\sigma=0.5$ and three different values of the solid volume
fraction: $\phi=0$ (solid line), $\phi=0.2$ (dashed line), and $\phi=0.4$ (dotted
line). \label{fig4}}
\end{figure}
\begin{figure}
\includegraphics[width=0.4 \columnwidth,angle=0]{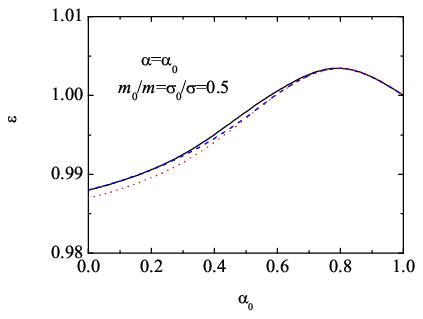}
\caption{(Color online) Plot of the Einstein ratio $\epsilon=D/T_0\mu$ versus the
coefficient of restitution $\alpha=\alpha_0$ in the case of the stochastic thermostat
for $d=2$, $m_0/m=\sigma_0/\sigma=0.5$ and three different values of the solid volume
fraction: $\phi=0$ (solid line), $\phi=0.2$ (dashed line), and $\phi=0.4$ (dotted
line). \label{fig5}}
\end{figure}
\begin{figure}
\includegraphics[width=0.4 \columnwidth,angle=0]{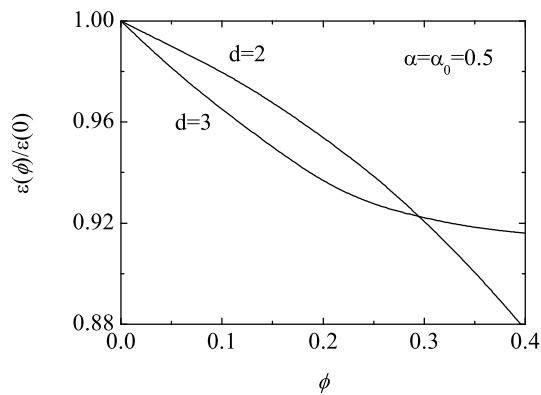}
\caption{Plot of the ratio $\epsilon(\phi)/\epsilon(0)$ versus the solid volume fraction $\phi$ for
$\alpha=\alpha_0=0.5$, $m_0/m=2$ and $\sigma_0/\sigma=0.25$  in the case of the Gaussian thermostat for spheres
($d=3$) and disks ($d=2$).   \label{fig6}}
\end{figure}

It is clear that $\epsilon$ becomes independent of the density $\phi$ when
$\sigma=\sigma_0$. As in the case of dilute gases \cite{G04}, Eqs.\ (\ref{24}) and
(\ref{25}) show that the violation of the Einstein relation in a heated moderately
dense granular gas is basically due to the departures of $f_0$ from its Maxwellian form
$f_{0,M}$. At this level of approximation (second Sonine approximation to the
coefficients $D$ and $\mu$), the deviation of $f_0$ from $f_{0,M}$ is only accounted
for by the fourth cumulant $c_0$. However, the coefficient $c_0$ depends on the solid
volume fraction $\phi$ through its dependence on the temperature ratio $T_0/T$. To
assess the influence of density on the Einstein ratio $\epsilon$, Fig.\ \ref{fig3}
shows a plot of $\epsilon$ versus the size ratio $\sigma_0/\sigma$ in the case of the
Gaussian thermostat for $d=3$ and $\alpha=\alpha_0=0.5$ when the impurities have the
same mass density as the gas particles [namely, $m_0/m=(\sigma_0/\sigma)^3$]. Three
different values of the solid volume fraction $\phi$ have been considered: $\phi=0$
(dilute gas), $\phi=0.2$ (moderate dense gas), and $\phi=0.4$ (quite dense gas). The
corresponding plot for the stochastic thermostat has not been included since the
deviation of $\epsilon$ from 1 is less than $1\%$ for all the cases analyzed. It is
apparent that the degree of violation of the Einstein relation is more important when
the impurities are lighter and/or smaller than the gas particles, especially for high
densities. To confirm this trend, the Einstein ratio has been plotted in Figs.\
\ref{fig4} and \ref{fig5} as a function of the (common) coefficient of restitution
$\alpha=\alpha_0$ for $d=2$, $m_0/m=\sigma_0/\sigma=0.5$, and for the same values of
$\phi$ as considered before. Figure \ref{fig4} shows the results obtained by using the
Gaussian thermostat and Fig.\ \ref{fig5} refers to the results obtained for the
stochastic thermostat. While $\epsilon$ is close to 1 in the case of the stochastic
thermostat for all the densities considered, significant deviations form unity are
observed for the Gaussian thermostat. In this latter case, it is apparent that the
degree of violation of the Einstein formula increases with the volume fraction and the
inelasticity.

This latter conclusion qualitatively agrees with the results obtained by Puglisi {\em
et al.} \cite{PBV07} from computer simulations since they observe a significant
violation of the Einstein formula when excluded volume effects and dissipation are
increased. However, at a quantitative level, the deviations observed by Puglisi {\em et
al.} \cite{PBV07} are larger than those found here [see Fig.\ 3 of Ref.\ \cite{PBV07}].
The quantitative disagreement between simulations \cite{PBV07} and the Enskog theory
results (\ref{24}) and (\ref{25}) could be due to two different and independent
reasons. First, as noted before in Section \ref{sec2}, the expression of the fourth
cumulant $c_0$ used here has been obtained by neglecting cumulants of higher order and
considering only linear terms in $c$ and $c_0$. On the other hand, according to the
previous results \cite{BP06,NBSG07} obtained in the free cooling case for a
monocomponent gas, the value of $c_0$ could change significantly for very strong
dissipation when cumulants of higher order were taken into account in the form of the
homogeneous distribution. In this context, the estimate of $c_0$ obtained in this paper
could be not reliable for this range of values of inelasticity and so, the quantitative
deviations of $\epsilon$ from $1$ observed in Figs.\ \ref{fig3} and \ref{fig4} for the
Gaussian thermostat when $\alpha_0\lesssim 0.5$ could be questionable. As a second
reason of discrepancy between simulations and theory, one could argue that the velocity
correlations (absent in the Enskog equation but present in computer simulations) play a
more important role than spatial correlations (excluded volume effects) in the
violation of the Einstein formula. In this case, one should correct the Enskog equation
by incorporating recollision events (``ring'' collisions) that take into account
multiparticle collisions.

Finally, since the results derived in this paper holds for a $d$-dimensional system, it
is interesting to investigate the influence of dimensionality on the violation of the
Einstein relation. To illustrate this effect, Fig.\ \ref{fig6} shows the dependence of
the ratio $\epsilon(\phi)/\epsilon(0)$ on the solid volume fraction $\phi$ for
$\alpha=\alpha_0=0.5$, $m_0/m=2$ and $\sigma_0/\sigma=0.25$  in the case of the
Gaussian thermostat. I have considered the physical cases of hard spheres ($d=3$) and
hard disks ($d=2$). Here, $\epsilon(0)$ corresponds to the value of the Einstein ratio
for a dilute gas. Although the qualitative dependence of the ratio
$\epsilon(\phi)/\epsilon(0)$ on $\phi$ is quite similar in both systems, we observe
that the violation of the Einstein ratio is stronger for $d=3$ than for $d=2$ for
moderate densities. However, this trend changes as density becomes larger.

\section{Conclusions}
\label{sec4}

In this paper I have analyzed the validity of the Einstein relation
$\epsilon=D/T_0\mu=1$ for driven moderately $d$-dimensional dense granular gases in the
framework of the Enskog equation. This work extends a previous study carried out by the
author \cite{G04} in the case of a dilute gas ($\phi=0$) of inelastic hard spheres
($d=3$). To achieve a NESS, two types of thermostats (external forces) have been
considered: (i) an ``anti-drag'' force proportional to the particle velocity (Gaussian
force), and (b) a stochastic force, which give frequent kicks to each particle between
collisions. The present work has been motivated by recent computer simulation results
by Puglisi {\em et al.} \cite{PBV07} where the spatial and velocity correlations
between the particles have shown to be the most important ingredient in a strong
violation of the Einstein relation. It is shown that $\epsilon \neq 1$, especially in
the case of the Gaussian thermostat when the impurity is lighter and/or smaller than
particles of the gas. As in the case of a dilute gas \cite{G04}, the violation of the
Einstein relation is connected to the strong non-Maxwellian behavior of the {\em
homogeneous} velocity distribution function of impurities, which is mainly measured
through its fourth cumulant $c_0$. The results also show that the deviation of the
Einstein ratio from 1 is more important as both the density and dissipation increase,
which is consistent with the observations made by Puglisi {\em et al.} \cite{PBV07}.
However, at a quantitative level, the deviations of the Einstein formula obtained here
from the Enskog equation are smaller than those found in computer simulations. As
discussed before, this quantitative disagreement between theory and simulation could be
due to (i) the possible lack of convergence of the Sonine polynomial expansion for the
reference Gaussian driven state, and/or  (ii) the influence of velocity correlations
which are absent in the Enskog description (molecular chaos hypothesis). With respect
to the first source of discrepancy, one perhaps should include cumulants of higher
order as well as nonlinear terms in $c$ and $c_0$ to get an accurate estimate of the
fourth cumulant $c_0$ for very small values of the coefficients of restitution $\alpha$
and $\alpha_0$. However, given that perhaps the absolute value of the higher order
cumulants increases with inelasticity, the Sonine expansion could be not relevant in
the sense that one would need an infinite number of Sonine coefficients to characterize
the reference state. In this case, a possible alternative would be the use of the
Direct Simulation Monte Carlo (DSMC) method \cite{B94} to numerically solve the Enskog
equation in the homogenous driven problem. Regarding the influence of velocity
correlations, the inclusion of this new ingredient in the Enskog collision operator
makes analytic calculations intractable since higher-order correlations must be
included in the evaluation of the collision integrals. This contrasts with the explicit
results reported in this paper, where the transport coefficients $D$ and $\mu$ have
been explicitly obtained in terms of the parameters of the system (masses, sizes and
coefficients of restitution).

Finally, it must be noted that the theoretical results derived here have been obtained
by considering the second Sonine approximation to the Chapman-Enskog solution. Exact
results can be obtained if one considers the inelastic Maxwell model (IMM) for a dilute
gas. This model has been widely used by several authors as a toy model to characterize
the influence of the inelasticity of collisions of the physical properties of the
granular fluids. The fact the the collision rate for the IMM  is velocity independent
allows one to exactly compute the transport coefficients of the system. In particular,
the coefficients $D$ and $\mu$ have been evaluated \cite{GA05} from the Chapman-Enskog
method for undriven systems. The extension of such calculations to driven systems is
straightforward. Thus, in the case of the Gaussian thermostat, one gets
\begin{equation}
\label{26} D=\frac{n_0T_0}{m_0}\left(\nu_D-\frac{1}{2}\zeta\right)^{-1}, \quad
\mu=\frac{n_0}{m_0}\left(\nu_D-\frac{1}{2}\zeta\right)^{-1},
\end{equation}
while
\begin{equation}
\label{27} D=\frac{n_0T_0}{m_0\nu_D}, \quad \mu=\frac{n_0}{m_0\nu_D}
\end{equation}
in the case of the stochastic thermostat. Here,
\begin{equation}
\label{28} \nu_D=\frac{w_0}{d}\frac{m}{m+m_0}(1+\alpha_0), \quad \zeta=\frac{w}{2d}(1-\alpha^2),
\end{equation}
where $w$ and $ w_0$ are effective collision frequencies of the model. According to Eqs.\ (\ref{26}) and
(\ref{27}), $\epsilon=D/T_0\mu=1$ for both thermostats so that, the Einstein relation holds for the {\em
inelastic} Maxwell model in any dimension. This conclusion agrees with previous independent results obtained for
$d=1$ \cite{SBL06,BBALMP05}, $d=2$ \cite{PBV07} and $d=3$ \cite{BSL08}.

\acknowledgments

This research has been supported by the Ministerio de Educaci\'on y Ciencia (Spain)
through grant No. FIS2007-60977, partially financed by FEDER funds, and by the Junta de
Extremadura (Spain) through Grant No. GRU08069.

\appendix
\section{Expressions of $\zeta_0$, $c_0$, $\nu_2$ and $\nu_4$}
\label{appA}

 The explicit expressions of the partial cooling rate $\zeta_0$, the kurtosis $c_0$ and the collision
frequencies $\nu_2$ and $\nu_4$ are displayed in this Appendix for an arbitrary number
of dimensions $d$. In order to get these expressions, we consider the leading Sonine
approximations (\ref{7.1}) for the granular gas distribution $f$ and (\ref{18.3}) for
the impurity distribution $f_0$. The cooling rate $\zeta_0$ can be obtained by
following the same mathematical steps as those used before in previous papers
\cite{G04,GM04}. The final expression can be written as
\begin{equation}
\label{a1} \zeta_0=\lambda_{00}+\lambda_{01}c_0+\lambda_{02}c,
\end{equation}
where
\begin{equation}
\label{a2} \lambda_{00}=\frac{4\pi^{(d-1)/2}}{d\Gamma\left(\frac{d}{2}\right)}\chi_0
n\overline{\sigma}^{d-1}v_{\text{th}} M \left(\frac{1+\theta}{\theta}\right)^{1/2}(1+\alpha_{0})
\left[1-\frac{M}{2}(1+\alpha_{0})(1+\theta)\right],
\end{equation}
\begin{equation}
\label{a3} \lambda_{01}=\frac{\pi^{(d-1)/2}}{8d\Gamma\left(\frac{d}{2}\right)}\chi_0
n\overline{\sigma}^{d-1}v_{\text{th}} M \frac{(1+\theta)^{-3/2}}{\theta^{1/2}}(1+\alpha_{0})
\left[2(3+4\theta)-3M(1+\alpha_{0})(1+\theta) \right],
\end{equation}
\begin{equation}
\label{a4} \lambda_{02}=-\frac{\pi^{(d-1)/2}}{8d\Gamma\left(\frac{d}{2}\right)}\chi_0
n\overline{\sigma}^{d-1}v_{\text{th}} M \left(\frac{1+\theta}{\theta}\right)^{-3/2}(1+\alpha_{0})
\left[2+3M(1+\alpha_{0})(1+\theta) \right].
\end{equation}
Here, $\theta=m_0T/mT_0$ and $M=m/(m+m_0)$.

In order to get the coefficient $c_0$, one substitutes Eqs.\ (\ref{7.1}) and
(\ref{18.3}) into the Enskog-Lorentz equation (\ref{16}), multiplies it by $v^4$ and
integrates over the velocity. After some algebra and neglecting nonlinear terms in $c$
and $c_0$, the result in the case of the Gaussian thermostat is
\begin{equation}
\label{a5}
c_0=-\frac{\lambda_{00}+\lambda_{02}c+\frac{2}{d(d+2)}M_0^{-2}\theta^2\left(\Omega_{00}+\Omega_{02}c\right)}
{\frac{1}{2}\lambda_{00}+\lambda_{01}+\frac{2}{d(d+2)} M_0^{-2}\theta^2\Omega_{01}},
\end{equation}
while
\begin{equation}
\label{a6}
c_0=-\frac{\lambda_{00}+\lambda_{02}c+\frac{2}{d(d+2)}M_0^{-2}\theta^2\left(\Omega_{00}+\Omega_{02}c\right)}
{\lambda_{01}+\frac{2}{d(d+2)}M_0^{-2}\theta^2\Omega_{01}}
\end{equation}
for the stochastic thermostat. In Eqs.\ (\ref{a5}) and (\ref{a6}), $M_0=m_0/(m+m_0)$ and the quantities
\begin{eqnarray}
\label{a7} \Omega_{00} &=&\frac{\pi^{(d-1)/2}}{\Gamma\left(\frac{d}{2}\right)}\chi_0
n\overline{\sigma}^{d-1}v_{\text{th}}M_0^2 M  \frac{\left( 1+\theta \right) ^{-1/2}}{\theta^{5/2}} \left(
1+\alpha_{0}\right) \nonumber\\
& & \times \left\{ -2\left[d+3+(d+2)\theta\right] +M\left( 1+\alpha _{0}\right) \left( 1+\theta \right)
\left( 11+ d+\frac{d^2+5d+6}{d+3} \theta \right)\right.\nonumber\\
& & \left. -8M^{2}\left( 1+\alpha _{0}\right) ^{2}\left( 1+\theta \right) ^{2} +2M^{3}\left( 1+\alpha
_{0}\right)^{3}\left( 1+\theta \right) ^{3}\right\} \;,
\end{eqnarray}
\begin{eqnarray}
\label{a8} \Omega_{01} &=&\frac{\pi^{(d-1)/2}}{16\Gamma\left(\frac{d}{2}\right)}\chi_0
n\overline{\sigma}^{d-1}v_{\text{th}}M_0^2 M  \frac{\left( 1+\theta \right) ^{-5/2}}{\theta^{5/2}} \left(
1+\alpha_{0}\right) \nonumber\\
& & \times
 \left\{-2\left[ 45+15d+(114+39d)\theta
+(88+32d)\theta^{2}+(16+8d)\theta^{3}\right] \right.  \nonumber \\
&&+3M\left( 1+\alpha _{0}\right) \left( 1+\theta\right) \left[ 55+5d+9(10+d)\theta +4(8+d)\theta
^{2}\right]\nonumber\\
& & \left. -24M^{2}\left( 1+\alpha _{0}\right) ^{2}\left( 1+\theta\right)^{2}\left( 5+4\theta\right)+30
M^{3}\left( 1+\alpha _{0}\right) ^{3}\left( 1+\theta\right) ^{3}\right\} \;,
\end{eqnarray}
\begin{eqnarray}
\label{a9} \Omega_{02}&=&\frac{\pi^{(d-1)/2}}{16\Gamma\left(\frac{d}{2}\right)}\chi_0
n\overline{\sigma}^{d-1}v_{\text{th}}M_0^2 M  \frac{\left( 1+\theta \right) ^{-5/2}}{\theta^{1/2}} \left(
1+\alpha_{0}\right) \nonumber\\
& & \times\left\{ 2\left[ d-1+(d+2)\theta\right] +3M\left( 1+\alpha
_{0}\right) \left( 1+\theta\right)\left[d-1+(d+2)\theta\right] \right.  \nonumber \\
& & \left. -24M^{2}\left( 1+\alpha _{0}\right)^{2}\left(1+\theta\right)^{2}+30M^{3} \left( 1+\alpha
_{0}\right)^{3}\left( 1+\theta\right)^{3}\right\}  \;,
\end{eqnarray}
have been introduced. Equations (\ref{a2})--(\ref{a4}) and (\ref{a7})--(\ref{a9}) are consistent with the
results \cite{G04,GD99} obtained for hard spheres ($d=3$). Once the coefficient $c_0$ is given in terms of
$\gamma=M_0/M \theta$, the parameters of the mixture and the solid volume fraction, the temperature ratio
$\gamma$ can be explicitly obtained by numerically solving the condition (\ref{17}) for the Gaussian thermostat
or the condition (\ref{18}) for the stochastic thermostat.

Finally, the collision frequencies $\nu_2$ and $\nu_4$ are given by
\begin{equation}
\label{a10} \nu_2=\frac{\pi^{(d-1)/2}}
{d\Gamma\left(\frac{d}{2}\right)}n\overline{\sigma}^{d-1}v_{\text{th}}\chi_{0} M(1+\alpha_{0})
[\theta(1+\theta)]^{-1/2},
\end{equation}
\begin{equation}
\label{a11} \nu_4=\frac{\pi^{(d-1)/2}}
{d(d+2)\Gamma\left(\frac{d}{2}\right)}n\overline{\sigma}^{d-1}v_{\text{th}}\chi_{0}
M(1+\alpha_{0})\left(\frac{\theta}{1+\theta}\right)^{3/2} \left[A-(d+2)\frac{1+\theta}{\theta} B\right],
\end{equation}
where
\begin{eqnarray}
\label{a12} A&=&
 2M^2\left(\frac{1+\theta}{\theta}\right)^{2}
\left(2\alpha_{0}^{2}-\frac{d+3}{2}\alpha_{0}+d+1\right)
\left[d+5+(d+2)\theta\right]\nonumber\\
& & -M(1+\theta) \left\{\beta\theta^{-2}[(d+5)+(d+2)\theta][(11+d)\alpha_{12}
-5d-7]\right.\nonumber\\
& & \left. -\theta^{-1}[20+d(15-7\alpha_{0})+d^2(1-\alpha_{0})-28\alpha_{0}] -(d+2)^2(1-\alpha_{0})\right\}
\nonumber\\
& & +3(d+3)\beta^2\theta^{-2}[d+5+(d+2)\theta]+ 2\beta\theta^{-1}[24+11d+d^2+(d+2)^2\theta]
\nonumber\\
& & +(d+2)\theta^{-1} [d+3+(d+8)\theta]-(d+2)(1+\theta)\theta^{-2} [d+3+(d+2)\theta],
\end{eqnarray}
\begin{eqnarray}
\label{a13} B&=& (d+2)(1+2\beta)+M(1+\theta)\left\{(d+2)(1-\alpha_{0})
-[(11+d)\alpha_{0}-5d-7]\beta\theta^{-1}\right\}\nonumber\\
& & +3(d+3)\beta^2\theta^{-1}+2M^2\left(2\alpha_{0}^{2}-\frac{d+3}{2}\alpha
_{0}+d+1\right)\theta^{-1}(1+\theta)^2\nonumber\\
& & - (d+2)\theta^{-1}(1+\theta).
\end{eqnarray}
Here, $\beta=M_0-M\theta=M_0(1-\gamma^{-1})$. For $d=3$, Eqs.\ (\ref{a10}) and (\ref{a11}) coincide with those
previously reported \cite{G04} for hard spheres.

\end{document}